\documentclass[11pt]{elsarticle}
\usepackage{amsmath,amssymb}
\usepackage[margin=2.5cm]{geometry}
\usepackage{graphics,graphicx}
\usepackage{dcolumn,bm}
\usepackage{psfrag}
\usepackage{color}
\newcommand{\be}{\begin{equation}}
\newcommand{\ee}{\end{equation}}

\usepackage{graphics,graphicx}
\journal{Physica A}

\begin{document}

\title{Kinetic exchange opinion dynamics for the battleground-states in the 2024 US presidential elections}

\author{Soumyajyoti Biswas\,$^{1,3}$, Parongama Sen\,$^{2}$, Bhargav Thota\,$^{3}$, Hemanth Kodali\,$^{3}$, I. Vinay Datta\,$^{3}$, K. Madhu Venkata Akash, $^{3}$}
\address{$^{1}$ Department of Physics, SRM University AP, Amaravati 522240, Andhra Pradesh, India.\\
  $^{2}$ Department of Physics, University of Calcutta, 92, A. P. C. Road, Kolkata 700009, India.\\
  $^{3}$ Department of Computer Science and Engineering, SRM University AP, Amaravati 522240, Andhra Pradesh, India.}

\date{\today}

\begin{abstract}
       The strongly polarizing political discourse in the U. S. implies that a small minority of the population, determining the outcome of the presidential elections in a few so called battleground-states, also determines the outcome of the overall election. Given the almost equal distributions of the electoral college members in the so-called blue and red states, the members elected from these battleground states would determine the election results. We build a kinetic exchange opinion model that takes into account the dynamical nature of the opinions of the individuals in the battleground states and the already determined core voters of the non-battleground states. In a fully connected graph, we consider the interaction among the population in the battleground states while the agents in the non-battleground states are assumed to have fixed opinions. We provide the analytical results and numerical simulations using realistic parameters from the opinion poll of the previous election's data. Counter-intuitively, a more noisy environment predicts a higher chance of the Democrats' win.       
\end{abstract}

\maketitle

\section{Introduction}
The U. S. presidential elections are historic events that have substantial influence on the geopolitical and economic status of the world in the subsequent years \cite{csis}. These are therefore, enthusiastically studied from various angles that are not only limited to political sciences (see e.g., \cite{shaw}), but also expanded into economics (see e.g., \cite{wang,erikson}), environment (see e.g., \cite{lancet}) and other disciplines \cite{geo}. 

The interests of mathematicians and scientists \cite{oup,galam,rmp} come from some of the stylized facts that are special, at least as a collection, to the US presidential elections. Firstly, the overwhelmingly bipolar character of the political discourse enables a simple quantification of the opinion values of the individual agents (voters) through an Ising-like variable (see e.g., \cite{galam1,snj,deff,toscani,hk,lccc,bcs}). The shifting of the winning majority over several election cycles indicate a near-critical status of the overall population, thereby opening the route of using the tools of critical phenomena for this complex social system. A near-critical status also indicates the dominance of a diverging correlation length that can suppress other details of the systems barring the symmetry of the local variable (here Ising-like) and the range of the interactions (here a fully connected graph for simplicity). Finally, the procedure of the election through the electoral college system, makes it an effective weighted course-graining \cite{cg1,cg2} that are widely studied in the critical phenomena \cite{stanley}.   

For the above mentioned reasons, and many other similar interests, mathematical quantification of opinion values and their evolutions have been studied using different dynamical rules (see e.g., \cite{oup,galam,rmp} for reviews) for several decades. The interests, among other things, are focused on the emergence of consensus or the lack thereof. The emergence of consensus, often cast in the context of appearance of a finite order parameter in systems undergoing phase transitions, are studied as critical behavior in interacting systems following simplistic dynamical evolution rules for its components (voters). The generic assumption, apart from the mathematical quantification of opinion values, is that the subtleties of human psychology would 'average-out' in the macroscopic measurable quantities, much like how the individual dynamical evolution of atoms in a gas at sufficiently high temperature could be cast as classical bouncing particles randomly moving in space, giving excellent estimates for the measures of thermodynamic variables. 

The kinetic theory of gas has been the motivation in formulating a class of opinion dynamics models \cite{toscani}, often identified as the kinetic exchange models that are applied in various contexts \cite{lccc,bcs,nuno1,nuno2,lima,kb}. These kinetic exchange opinion models are known to show a phase transition to consensus from random opinion states, typically in the Ising universality class \cite{sudip,solomon}. The fluctuations of the opinion states near the critical point result in several interesting phenomena such as the minority-win in the US presidential election \cite{cg1,cg2}, where the candidate winning more popular votes might not end up winning less electoral college votes (2016 election being the most recent example) and also a long waiting time in emergence of consensus due to formation of strongly oriented opposing domains of opinion values (Brexit being the latest example) \cite{brexit}. Several of these phenomena could be compared with available data at a high level yielding satisfactory broad similarities. 

In this work, we examine the 2024 U. S. presidential elections focusing on the seven perceived battleground states, which have known history of voting for candidates of either parties and are presently almost evenly divided in opinion polls. We consider the interactions of the individuals in these states through the kinetic exchange opinion models. Additionally, we also consider the opinions of the individuals of the so called blue and red states, but assume that while they can still influence other undecided voters, they do not change their own opinion values. We present both analytical calculations and numerical simulations of this system. As we shall see, given the parameters estimated from earlier elections and current opinion polls, an environment of higher 'noise/discontent' would favor, rather counter-intuitively, a Democratic victory.  

\section{Model}
We use a kinetic exchange opinion model, introduced in \cite{bcs}, for the battleground states: Arizona, Georgia, Pennsylvania, Nevada, North Carolina, Michigan and Wisconsin. Particularly, the opinion value of the $i$-th agent is written as $o_i(t)$. At each interaction, another agent is randomly chosen (binary interaction in a fully connected graph), and the evolution is as follows:
\begin{equation}
    o_i(t+1)=o_i(t)+\mu o_j(t),
\end{equation}
where $\mu$ can take values $-1$ or $+1$ with probabilities $p$ and $1-p$ respectively and represent either a negative or a positive interaction between the two selected agents. Along with the dynamical equation, ceilings at $\pm 1$ are enforced to keep the opinion values bounded.  However, along with the individuals in the battleground states, there are individuals in the so called blue and red states, for which we assume that they are committed voters and while they can influence an undecided voter in the battleground state, they do not change their own votes i.e., their opinion values are fixed at either $+1$ (for Democrats) or $-1$ (for Republicans). In other words, they are 'inflexibles' in the similar spirit as considered in \cite{inflex,inflex1}. Note that during the dynamics, given the eolution rules, the allowed opinion values of individuals are only $\pm 1$ and $0$.

We denote by $r$ the fraction of the population of the  undecided voters that can still modify their opinions. We assume these are the voters in the battle ground states. 
Considering the voting populations of the battleground states, it is approximately 20\% of the total number of voters ($r\approx 0.2$). Therefore, when the opinion of an agent is updated, 20\% of time it interacts with another agent in the battleground states. For the remaining 80\% time, assuming uniform and complete connectivity for simplicity, the randomly chosen agent interacts with one of the inflexible agents from the non-battle ground states. Now, we assume $f_d$ to be the fraction of the inflexible voters who are committed Democrats (having opinion values $+1$) and $1-f_d$ fractions are committed Republicans (having opinion values $-1$). In practice there is a small fraction of voters, often committed, outside of the two parties, but we do not consider it here. The dynamical equation, of course, allows for zero opinion values that can represent third party voters and neutral individuals. One time step in the model is represented by $N$ randomly chosen updates. In this work, we are only interested in the steady-state properties of the results.

\begin{figure}[ht]
\centering
\includegraphics[width=14.0cm, keepaspectratio]{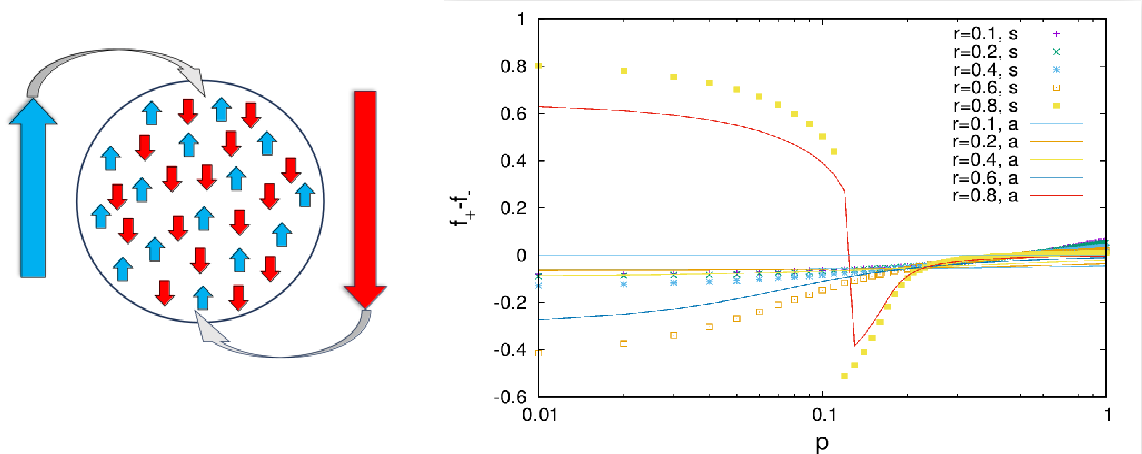}  
\caption{The left figure is a schematic diagram of the model. The large arrows denote the inflexible Democrats and Republicans in the non-battle ground states. The arrows within the circles represent the undecided voters (the voters in the battle ground states). The right side figure is a comparison between the numerical simulations of the model (points) and the corresponding analytical results of the (signed) order parameter. The initial conditions are such that the system (flexible part) starts with all up and then evolves to a steady state. Depending upon the values of $p$, the steady state can be a Democrat or a Republican majority (in the battle ground states).}
\label{fig1}
\end{figure}
\section{Results}
Given the fully connected nature of the model, a mean-field theory calculation can be done for the model dynamics. Below, we first present the calculations and then go on to verify the order parameter and some other quantities using numerical simulations. 

\subsection{Mean field theory}

Let a fraction $1-r$ of the population be  committed voters (inflexibles) out of  which $f_d$ is the fraction voting for Democrats  and their opinion state is $+1$. Hence $(1-f_d)$  is the fraction out of the inflexibles who vote for the Republicans and we assume the opinion to be $-1$. We are not considering neutral (or third party) committed voters, which is anyway a small fraction.

Out of the  voters in the battleground states  (fraction $r$ of the population), let $f_{\pm}$ 
be the fraction voting for Democrats/Republicans  and the undecided fraction with zero state is $f_0$.                
The mean field rate equations can then be written down as
\begin{eqnarray}
\frac{df_{+}}{dt} &= &-p(rf_{+} + (1-r) f_d)f_{+}  - (1-p)(rf_{-} + (1-r)(1- f_d)) f_{+} \nonumber \\
& & + p(f_{-} + (1-r)(1-f_d))f_0 + (1-p)(rf_{+} + (1-r)f_d)f_0
\end{eqnarray}
 \begin{eqnarray}
\frac{df_{-}}{dt} &= &-p(rf_{-} + (1-r) (1-f_d))f_{-}  - (1-p)(rf_{+} + (1-r)f_d) f_{-} \nonumber 
\\
& & + p(f_{+} + (1-r)f_d)f_0 + (1-p)(rf_{-} + (1-r)(1-f_d))f_0
\end{eqnarray}

The steady state values of the order parameter are then obtained by numerically solving the above equations from arbitrary initial states and studied as a function of $p$ for fixed values of $r$ and $f_d$. 
\begin{figure*}[ht]
\centering
\includegraphics[width=8.0cm, keepaspectratio]{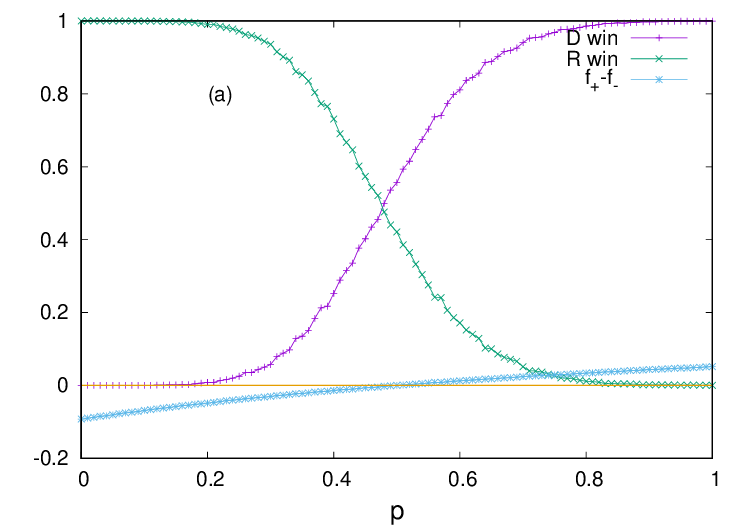}  
\includegraphics[width=8.0cm, keepaspectratio]{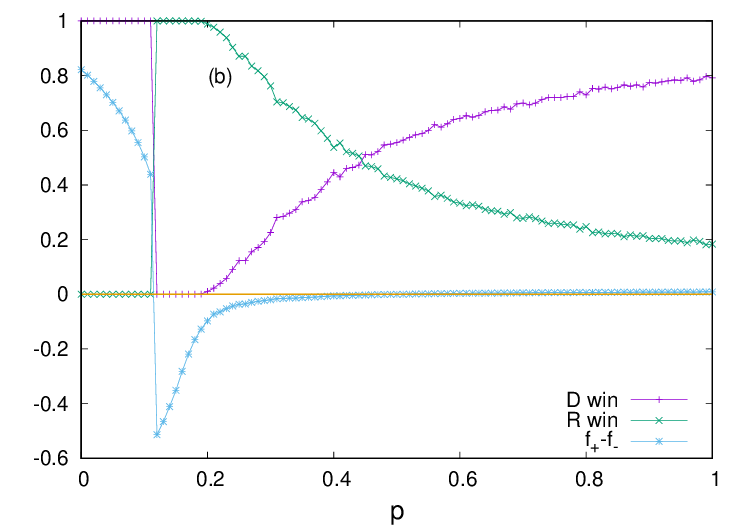}
\caption{The probabilities of an over all Democratic win (D-win) or a Republican win (R-win) and the difference of fraction of Democrats ($f_+$) and Republican ($f_-$) voters within the battle ground states. (a) Here the fraction of flexible (battle ground state) population is taken as $r=0.2$. Initial condition is $f_+=1$, and $f_d=0.471$ (see text). For low values of $p$, the ordering of the battle ground voters follow the ordering of the inflexible voters (Republican majority). But for $p>0.5$, the order in the battle ground states are opposite to the order of the non-battle ground state. The probability of Democrats winning (over all) increases. (b) The same is done for $r=0.8$. Here, due to the initial condition $f_+(t=0)=1$, even for low $p$ a Democratic majority is sustained within the battle ground state. But for an intermediate $p$, the majority in the battle ground states swicthes abruptly to Republican majority. It then again switches back to a small Democratic majority for $p>0.5$.}
\label{fig2}
\end{figure*}
\subsection{Numerical simulations}
For the numerical simulations of the models, we take the system size to be $N=3421$ or its integral multiples, to maintain proportionality of the voting populations of the seven battleground states (see Table 1). Between these states, the total electoral college vote is 93. These 93 members, each group assigned to the winner of the corresponding states, would decide the final tally of the electoral college and hence the election, since the other members of the electoral college are expected to be almost evenly split: 226-219 in favor of Democrats. 

\begin{figure*}[ht]
\centering
\includegraphics[width=14.0cm, keepaspectratio]{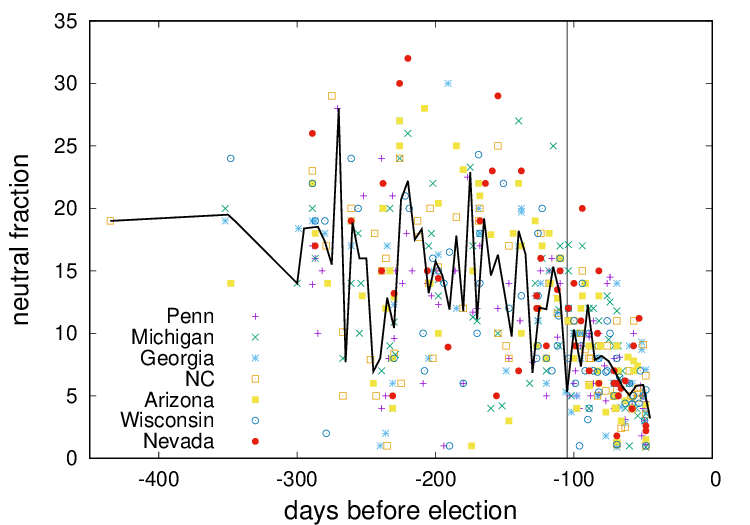}  
\caption{The fraction of the voters, follow surveys \cite{survey}, in the battle ground states who do not vote for Democrats or Republicans. This so called neutral fraction decreases as the election approaches. The vertical line denotes the time when Biden withdrew from the race in favor of Harris.}
\label{fig3}
\end{figure*}

There is another aspect of the electoral college which becomes important in these tightly fought elections. Due to the winner-takes-it-all rule for most of the states (in all of the seven states mentioned), it might happen that the candidate winning most number of popular votes might not end up winning most number of the electoral college votes, the two latest examples being 2016 and 2000 elections.

In Fig. \ref{fig1}(b) a comparison of the numerical simulations (points) and analytical calculations (solid lines) are shown. These are for an initial condition where $f_+>f_-$ at $t=0$. This shows that when $r$ is relatively large, an order opposite to the order of the inflexible voters could be sustained for smaller values of $p$. As $p$ is increased, this abruptly switches over to the ordering direction of the inflexible voters. 

The ordering direction of inflexible voters (i.e., the voters in the non-battle ground states) are towards a Republican majority, with $f_d=0.471$ of Democratic voters and rest being Republicans. This value is taken in line with the voting pattern of the 2020 election cycle \cite{2020_data} of the non-battle ground states (all states and D.C. without the 7 states listed in Table 1). 

When $p>1/2$, an anti-ferromagnetic order is preferred. Of course, in this mean field scenario, there is no spatial order. But this is manifested by an ordering of the flexible voters (voters in the battle ground states) building an order in the direction opposite to that present among the inflexible voters. Recall that high $p$ value imply highly negative interactions, hence pushing an ordering opposite to the predominant order among non-battle ground states.

In Fig. \ref{fig2}, we show the overall probabilities for a Democratic and Republican win. This is estimated in the following way: For the non-battle ground states, the electoral college is split as 226-219 in favor of the Democrats (see \cite{ec_data}). The 93 battle ground states electorates ar assigned based on the majority within the groups representing the populations of each state within the $N$ agents. These are just randomly assigned individuals for each state, proportional in numbers to the voting populations in each of the states (see Table 1). That is why $N$ is always taken as an integral multiple of 3421 (total voting population in the battle ground states in the ratio of 100 to 1 million). Finally, after reaching the steady state, the party with the most number of electorate (electoral college representatives) wins. Note that this is not just following the majority of the battle ground states. But in practice, that is nearly the case given the almost even split of the other states.  The two figures in Fig. \ref{fig2}, are for $r=0.2$ (a) and $r=0.8$ (b), respectively. The more realistic scenario is $r=0.2$.

\begin{figure*}[ht]
\centering
\includegraphics[width=8.0cm, keepaspectratio]{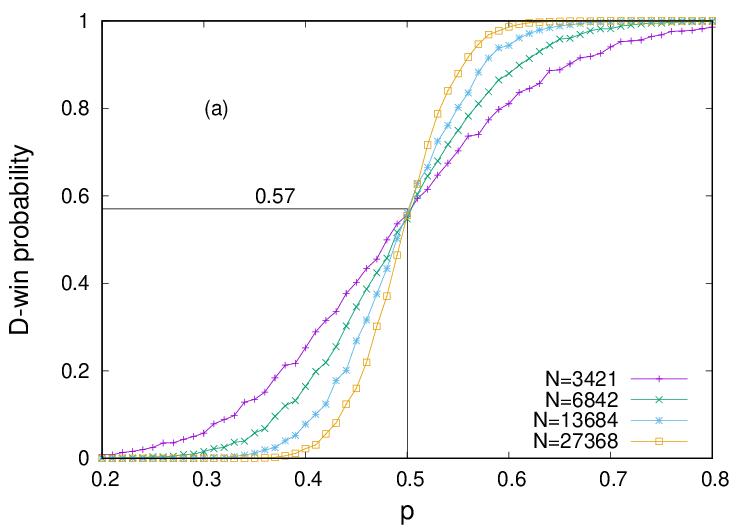}  
\includegraphics[width=8.0cm, keepaspectratio]{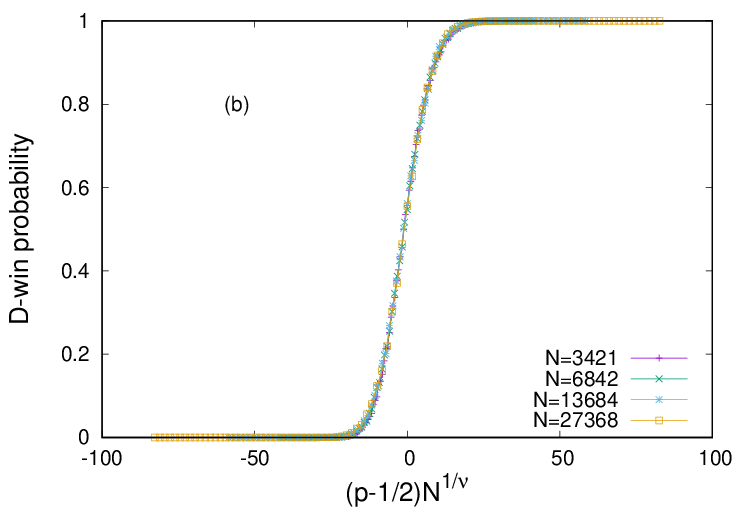}
\caption{ (a) The over all winning probabilities for Democrats and Republicans when $r=0.2$ and $f_d=0.471$ for different system sizes. For $p=1/2$, the Democratic win probability is $0.57$. The finite size collapse is shown in (b) with $\nu=1/2$.}
\label{fig4}
\end{figure*}

We wish to highlight that as the election approaches, the number of neutral voters decreases, which is also a signature of negative interactions. We argue that more voters are now willing to take a side, as they do not like the predominant outcome. In Fig. \ref{fig3}, we plot the percentages of neutral voters (specifically, those who did not commit to Democrats or Republicans) as the election approaches. The vertical line denotes the time when Biden withdrew from the race in favor of Harris. Even from the time before that, the neutral fraction was decreasing, as indicated by the solid average line. 

Finally, in Fig. \ref{fig4}, we do a finite size scaling analysis of the Democratic winning probability. Note that for all simulations, $r=0.2$ and $f_d=0.471$ here. We argue that as negative interaction grows, a Democratic win becomes more likely. At the least, when $p=1/2$, the Democratic win probability is about 57\%.

\begin{table}[h]
    \centering
    \caption{The number of voters, electoral college seats and net leading candidates in the battle ground states (+ for Democrats and - for Republicans) \cite{ec_data}}
    \begin{tabular}{|l|c|c|c|}
        \hline
        \textbf{State name} & \textbf{Electoral College} & \textbf{Registered Voters (millions)} & \textbf{Net Opinion} \\
        \hline
        Arizona & 11 & 3.88 & -2.2 \\
        \hline
        Georgia & 16 & 7.48 & -2.4 \\
        \hline
        Pennsylvania & 19 & 7.34 & -0.6 \\
        \hline
        Nevada & 6 & 1.45 & -0.5 \\
        \hline
        North Carolina & 16 & 5.16 & -1.0 \\
        \hline
        Michigan & 15 & 5.51 & 0.5 \\
        \hline
        Wisconsin & 10 & 3.39 & -0.6 \\
        \hline
    \end{tabular}
\end{table}


\section{Discussions and conclusions}
The political discourse in the last couple of decades, including in the U.S. has never been more polarized since the Cold War. Outside the geopolitical complications among the countries, the debates and discussions within the countries, including the U.S. has become highly polarized. The tightly fought Presidential elections are manifestations of that. 

Given the predominantly two-party scenario, and the electoral college system, it is the so called battle ground states that are in the focus of this election. We have applied a kinetic exchange opinion model for these battle ground sates, with as far realistic parameter values as possible, keeping in mind the already decided voters of the non-battle ground states. We show that the combined effect of having a Republican majority (of popular votes) in the non-battle ground states and a presumed highly negative interaction among the undecided voters, an order opposite to the prevailing order in the non-battle ground states is likely to form within the battle ground states. This in-turn favors a Democratic win overall, given the electoral college system. 

Of course, there are many assumptions in these processes including the model, which is just binary exchange and on a complete graph. Also, the voter fractions in the non-battle ground states are taken from the 2020 election cycles, and are assumed to be inflexible. Similarly, the entire population of the battle ground states are not necessarily flexible voters. Nevertheless, the mechanism of formation of an opposite order within the flexible population than what is present among the inflexible population, is independent of the exact values used, but is a result of high negative interaction i.e., $p>1/2$. 

In conclusion, with the present assumptions of inflexible voters within the non-battle ground states and flexible voters within the battle ground states, following a kinetic opinion exchange dynamics, a higher chance of at least 57\% of a Democratic victory is predicted in this study.

\end{document}